\documentclass[journal]{IEEEtran}
\usepackage{amssymb,epsfig,color,cite,amsmath,amsfonts,mathrsfs,algorithm,bm,algorithmicx}
\usepackage{float,balance}
\usepackage{color}
\usepackage{caption}
\usepackage{graphicx}
\newtheorem{lem}{Lemma}[section]
\newtheorem{thm}{Theorem}[section]

\newtheorem{rem}{Remark}[section]

\newtheorem{asm}{Assumption}[section]

\newcommand{\bed}{\begin{displaymath}}
\newcommand{\eed}{\end{displaymath}}
\newcommand{\bea}{\bed\begin{array}{rl}}
\newcommand{\eea}{\end{array}\eed}
\newcommand{\beq}[1]{\begin{equation} \label{#1}}
\newcommand{\eeq}{\end{equation}}
\newcommand{\barray}{\begin{array}{ll}}
\newcommand{\earray}{\end{array}}

\def\one{{\hbox{1{\kern -0.42em}1}}}

\begin{document}

\title{Convergence of a Distributed Least Squares}

\author{Siyu Xie,~\IEEEmembership{}
        Yaqi Zhang,~\IEEEmembership{}
        and Lei Guo,~\IEEEmembership{Fellow, IEEE}
\thanks{S. Y. Xie is with the Department of Electrical and Computer Engineering, Wayne State University, Detroit, MI 48202, USA. Email: syxie@wayne.edu.

Y. Q. Zhang and L. Guo are with Key Laboratory of Systems and Control, Academy of Mathematics and Systems Science, Chinese Academy of Sciences, Beijing 100190, P. R. China. They are also with School of Mathematical Science, University of Chinese Academy of Sciences, Beijing 100049, P. R. China. Email : \emph{zhangyq@amss.ac.cn., lguo@amss.ac.cn.}}
\thanks{This work was supported by the National Natural Science Foundation of China under grant 11688101.}
}

\markboth{IEEE Transactions on Automatic Control}{Xie and Guo: Convergence of a Distributed Least Squares}

\maketitle

\begin{abstract}
In this paper, we consider a least-squares (LS)-based distributed algorithm build on a sensor network to estimate an unknown parameter vector of a dynamical system, where each sensor in the network has  partial information only but is allowed to communicate with its neighbors. Our main task is to generalize the well-known theoretical results on the traditional LS to the current distributed case by establishing both the upper bound of the accumulated regrets of the adaptive predictor and the convergence of the distributed LS estimator, with the following key features compared with the existing literature on distributed estimation: Firstly, our theory does not need the previously imposed independence, stationarity or Gaussian property on the system signals, and hence is applicable to stochastic systems with feedback. Secondly, the cooperative excitation condition introduced and used in this paper for the convergence of the distributed LS estimate is the weakest possible one, which shows that even if any individual sensor cannot estimate the unknown parameter by the traditional LS, the whole network can still fulfill the estimation task by the distributed LS. \end{abstract}


\IEEEpeerreviewmaketitle

\section{Introduction}


Distributed estimation over sensor networks has received increasing research attention recently, and has been studied and used in many areas widely, e.g., collaborative spectral sensing in cognitive radio systems, target localization in biological networks, environmental monitoring, military surveillance, and so on \cite{Sayed2013}. Note that different cooperation strategies will lead to different distributed estimation algorithms, for examples, incremental least mean squres (LMS) \cite{Sayed2014b}, consensus LMS \cite{Xie2018Auto, Xie2018SIAM}, diffusion LMS \cite{Xie2018TAC, Har2019}, incremental LS \cite{Sayed2006}, consensus LS \cite{Mateos2012}, diffusion LS \cite{Xiao2006, Cattivelli2008, Bertrand2011, Ara2014, Vah2017, Ras2019, Yu2019}, and distributed Kalman filter (KF) \cite{Carli2008, Bat2015, Liu2018, Das2017}. In our recent work (see e.g. \cite{Xie2018Auto, Xie2018SIAM, Xie2018TAC}), we have given the stability and performance results for the consensus and diffusion LMS filters, without imposing the usual independence and stationarity assumptions for the system signals.

Note that when the unknown parameter is time-invariant, the LS algorithm may generate more accurate estimates in the transient phase and have faster convergence speed compared with LMS algorithm. This is one of the main motivations for us to consider the LS-based distributed estimation algorithm in this paper. Another reason for us to study this problem is that the existing convergence theory in the literature \cite{Sayed2006, Mateos2012, Xiao2006, Cattivelli2008, Bertrand2011, Ara2014, Vah2017, Ras2019, Yu2019} can hardly be applied to non-independent and non-stationary signals coming from practical complex systems where feedback loops exist inevitably. 

Fortunately, in the traditional single sensor case, there is a vast literature on the convergence theory of the classical LS, which is indeed applicable to stochastic systems with feedback. In fact, motivated by the need to establish a rigorous theory for the well-known LS-based self-tuning regulators proposed by $\mathring{\text{A}}$str\" om and Wittenmark \cite{Astrom1973} in stochastic adaptive control, the convergence study of LS with possible stochastic feedback signals had received a great deal of attention in the literature, see e.g., \cite{Ljung1976, Moore1978, Chen1982, Lai1982, Lai1986, Chen1986, Chen1991, Guo1991, Guo1995}. At the same time, much effort had also been devoted to stochastic adaptive control, see e.g, \cite{Ljung1977, Goodwin1981, Kumar1990}. Among the many significant contributions in this direction, here we only mention that Lai and Wei \cite{Lai1982} established a celebrated convergence result under a weakest possible decaying excitation condition on the system signals, and Guo and Chen \cite{Guo1991} and Guo \cite{Guo1995} finally resolved the longstanding problem concerning the global stability and convergence of the LS-based self-tuning regulators.

In this paper, we will provide a theoretical analysis for a distributed LS algorithm of diffusion type \cite{Bat2015, Liu2018}, where the diffusion strategy is designed via the so called covariance intersection fusion rule \cite{Julier1997, Chen2002}. In such a diffusion strategy, each node is only allowed to communicate with its neighbors, and both the estimates of the unknown parameter and the inverse of the covariance matrices are diffused between neighboring nodes. We will generalize the well-known convergence results on the classical LS by establishing both the upper bound of the accumulated regrets of the adaptive predictor and the convergence of the distributed LS estimator, with the following key features compared with the related results in the existing literature:
\begin{itemize}
\item Our theory does not need the usually assumed independence, stationarity or Gaussian property on the system signals, and hence does not exclude the applications of the theory to stochastic feedback systems.
\item Our theory for the convergence of the distributed LS is established under a weakest possible cooperative excitation condition which is a natural extension of the single sensor case. The cooperative excitation condition introduced in this paper implies that even if any individual sensor is not able to estimate the unknown parameter, the distributed LS can still accomplish the estimation task.
\end{itemize}

The rest of the paper is organized as follows. In Section II, we present some preliminaries on notations and graph theory, the observation model, and the distributed LS algorithm studied in the paper. The main results are stated in Section III. In Section IV, we provide the proofs of the main results. Finally, some concluding remarks are given in Section V.

\section{Problem Formulation}

\subsection{Basic Notations}

Let $A\in\mathbb{R}^{n\times n}$ and $B\in\mathbb{R}^{n\times n}$ be two symmetric matrices with real entries, then $A\geq B (A>B)$ means $A-B$ is a positive semidefinite (definite) matrix. Also, let $\lambda_{\max}\{\cdot\}$ and $\lambda_{\min}\{\cdot\}$ denote the largest and the smallest eigenvalues of the corresponding matrix, respectively. For any matrix $X\in\mathbb{R}^{m\times n}$, $\parallel X\parallel$ denotes the operator norm induced by the Euclidean norm, i.e., $(\lambda_{\max}\{XX^{\top}\})^{\frac{1}{2}}$ , where $(\cdot)^{\top}$ denotes the transpose operator. We use $\mathbb{E}[\cdot]$ to denote the mathematical expectation operator, and $\mathbb{E}[\cdot|\mathcal{F}_{k}]$ to denote the conditional mathematical expectation operator, where $\{\mathcal{F}_{k}\}$ is a sequence of nondecreasing $\sigma$-algebras\cite{Chow1978}. We also use $\log (\cdot)$ to denote the natural logarithm function, and $\text{Tr}(\cdot)$ to denote the trace of the corresponding matrix. Through out the paper, $\vert\cdot\vert$ denotes the determinant of the corresponding matrix, which should not be confused with the absolute value of a scalar from the context.

Let $\{A_{k},k\geq 0\}$ be a matrix sequence and $\{b_{k},k\geq 0\}$ be a positive scalar sequence. Then by $A_{k}=O(b_{k})$ we mean that there exists a constant $M>0$ such that $\|A_{k}\|\leq Mb_{k}, \forall k\geq 0$, and by $A_{k}=o(b_{k})$ we mean that $\lim\limits_{k\to\infty}\|A_{k}\|/b_{k}=0$.

\subsection{Graph Theory}

As usual, let the communication structure among sensors be represented by an undirected weighted graph $\mathcal{G}=(\mathcal{V},\mathcal{E},\mathcal{A})$, where $\mathcal{V}=\{1, 2,......, n\}$ is the set of sensors and $\mathcal{E}\subseteq \mathcal{V}\times \mathcal{V}$ is the set of edges. The structure of the graph $\mathcal{G}$ is described
by $\mathcal{A}=\{a_{ij}\}_{n\times n}$ which is called the weighted adjacency matrix, where $a_{ij}>0$ if $(i, j)\in\mathcal{E}$ and $a_{ij}=0$ otherwise. In this paper, we assume that the elements of the weighted matrix $\mathcal{A}$ satisfy $a_{ij}=a_{ji}, \forall i,j=1,\dots,n$, and $\sum_{j=1}^{n}a_{ij}=1,\forall i=1,\dots,n$.  Thus the matrix $\mathcal{A}$ is symmetric and doubly stochastic\footnote{A matrix is called doubly stochastic, if all elements are nonnegative, both the sum of each row and the sum of each column equal to 1.}. 

A path of length $\ell$ in the graph $\mathcal{G}$ is a sequence of nodes $\{i_{1},\dots,i_{\ell}\}$ subject to $(i_{j},i_{j+1})\in\mathcal{E}$, for $1\leq j\leq\ell-1$. The maximum value of the distances between any two nodes in the graph $\mathcal{G}$ is called the diameter of $\mathcal{G}$. Here in this paper, we assume that the graph is connected, and denote the diameter of the graph $\mathcal{G}$ as $D_{\mathcal{G}}$. Then $1\leq D_{\mathcal{G}}<\infty$ holds. The set of neighbors of the sensor $i$ is denoted as $
\mathcal{N}_{i}=\{j\in V | (j,i)\in \mathcal{E}\}$, and the sensor $i$ can only share information with its neighboring sensors from $\mathcal{N}_{i}$.

\subsection{Observation Model}

Let us consider a sensor network consisting of $n$ sensors. Assume that at each time instant $k$, each sensor $i\in\{1,\dots,n\}$ in the sensor network receives a noisy scalar measurement $y_{k+1,i}$ and an $m$-dimensional regressor $\bm{\varphi}_{k,i}\in\mathbb{R}^{m}$. They are related by a typical linear stochastic regression model
\begin{equation}\label{model}
y_{k+1,i}=\bm{\varphi}_{k,i}^{\top}\bm{\theta}+w_{k+1,i}, ~~~~k\geq 0,
\end{equation}
where $w_{k+1,i}$ is a random noise process, and $\bm{\theta}\in\mathbb{R}^{m}$ is an unknown parameter vector which needs to be estimated. We assume that at any time instant $k\geq 1$, each sensor $i$ uses both the observations $y_{j+1,i}$ and the regressors $\bm{\varphi}_{j,i} (j\leq k)$ to estimate the unknown parameter $\bm{\theta}$.

\subsection{Distributed LS Algorithm}

We now consider the following basic class of distributed LS algorithms of diffusion type:

\begin{algorithm}\label{LMS2}
\caption{A Distributed LS algorithm}
For any given sensor $i\in\{1\dots,n\}$, begin with an initial estimate $\bm{\theta}_{0,i}\in\mathbb{R}^{m}$, and an initial positive definite matrix $P_{0,i}\in\mathbb{R}^{m\times m}$. The algorithm is recursively defined at any time instant $k\geq 0$ as follows:
\begin{algorithmic}[1]
\State Adapt (generate $\bar{\bm{\theta}}_{k+1,i}$ and $\bar{P}_{k+1,i}$ on the bases of $\bm{\theta}_{k,i}, P_{k,i}$ and $\bm{\varphi}_{k,i}, y_{k+1,i}$):
\begin{equation}\label{DLS1}
\bar{\bm{\theta}}_{k+1,i}=\bm{\theta}_{k,i}+b_{k,i}P_{k,i}\bm{\varphi}_{k,i}(y_{k+1,i}-\bm{\varphi}_{k,i}^{\top}\bm{\theta}_{k,i}),
\end{equation}
\begin{equation}\label{DLS2}
\bar{P}_{k+1,i}=P_{k,i}-b_{k,i}P_{k,i}\bm{\varphi}_{k,i}\bm{\varphi}_{k,i}^{\top}P_{k,i},
\end{equation}
\begin{equation}\label{DLS3}
b_{k,i}=(1+\bm{\varphi}_{k,i}^{\top}P_{k,i}\bm{\varphi}_{k,i})^{-1},
\end{equation}

\State Combine (generate $P_{k+1,i}^{-1}$ and $\bm{\theta}_{k+1,i}$ by a convex combination of $\bar{P}_{k+1,j}^{-1}$ and $\bar{\bm{\theta}}_{k+1,j}$):
\begin{equation}\label{DLS4}
P_{k+1,i}^{-1}=\sum_{j\in\mathcal{N}_{i}}a_{ji}\bar{P}_{k+1,j}^{-1},
\end{equation}
\begin{equation}\label{DLS5}
\bm{\theta}_{k+1,i}=P_{k+1,i}\sum_{j\in\mathcal{N}_{i}}a_{ji}\bar{P}_{k+1,j}^{-1}\bar{\bm{\theta}}_{k+1,j}.
\end{equation}
\end{algorithmic}
\end{algorithm}


\begin{rem}
When $\mathcal{A}=I_{n}$, the  distributed LS will degenerate to the classical LS at any sensor $i$. Note that for stochastic gradient-based \cite{Chen1991} and LMS-based \cite{Guo} distributed estimation algorithms, the communication complexity may be reduced. However, for those algorithms, the estimation error either converges slowly to zero or does not converge to zero at all. Therefore, there is a tradeoff between the complexity and the convergence rate of the distributed estimation algorithms. Moreover, the convergence rate would be ``optimal'' when $\bar{P}_{k,i}$ is chosen to be the form in the paper. Furthermore, some existing methods may be used to reduce the communication complexity and to make the algorithm suitable for higher dimensional signals, for examples, event-driven methods \cite{Zhong2010}, partial diffusion methods \cite{Ara2014, Vah2017}, and compressed methods \cite{Xie2020} and so on. 
\end{rem}

\section{The Main Results}

\subsection{Some Preliminaries}

For the theoretical analysis, we need some standard assumptions on noise processes, regressors, and network topology.

\begin{asm}\label{asm1}
For each $i\in\{1,\dots,n\}$, the noise sequence $\{w_{k,i},\mathcal{F}_{k}\}$ is a martingale difference (where $\{\mathcal{F}_{k}\}$ is a sequence of nondecreasing $\sigma$-algebras), and there exists a constant $\beta>2$ such that $
\sup_{k\geq 0}\mathbb{E}[\vert w_{k+1,i}\vert^{\beta}\vert\mathcal{F}_{k}]<\infty, \ \hbox{a.s.}$
\end{asm}

\begin{asm}\label{asm2}
For each $i\in\{1,\dots,n\}$, the regressor sequence $\{\bm{\varphi}_{k,i},\mathcal{F}_{k}\}$ is an adapted sequence.
\end{asm}

\begin{asm}\label{asm3}
The graph $\mathcal{G}$ is connected.
\end{asm}

\begin{rem}
From \emph{Lemma 8.1.2} in \cite{Godsil2014}, it is not difficult to see that for any two nodes $i$ and $j$, there exists a path from $i$ to $j$ with length not larger than $\ell$ if and only if the $(i,j)$th entry of the matrix $\mathcal{A}^{\ell}$ is positive. From this, it is easy to see that each entry of the matrix $\mathcal{A}^{\ell}$ will be positive when $\ell$ is not smaller than the diameter of the graph $\mathcal{G}$, i.e., $D_{\mathcal{G}}$, see also \cite{Liu2018}.
\end{rem}

\subsection{Theoretical Results}

Here we first introduce the following notations:
$$
\begin{aligned}
&\bm{Y}_{k}\overset{\triangle}{=}\text{col}\{y_{k,1},\dots,y_{k,n}\},~~\bm{\Phi}_{k}\overset{\triangle}{=}\text{diag}\{\bm{\varphi}_{k,1},\dots,\bm{\varphi}_{k,n}\},\\
&\bm{W}_{k}\overset{\triangle}{=}\text{col}\{w_{k,1},\dots,w_{k,n}\},~~\bm{\Theta}\overset{\triangle}{=}\text{col}\{\underbrace{\bm{\theta},\dots,\bm{\theta}}_{n}\},\\
&\bm{\Theta}_{k}\overset{\triangle}{=}\text{col}\{\bm{\theta}_{k,1},\dots,\bm{\theta}_{k,n}\},~~\bar{\bm{\Theta}}_{k}\overset{\triangle}{=}\text{col}\{\bar{\bm{\theta}}_{k,1},\dots,\bar{\bm{\theta}}_{k,n}\},\\
&\widetilde{\bm{\Theta}}_{k}\overset{\triangle}{=}\text{col}\{\widetilde{\bm{\theta}}_{k,1},\dots,\widetilde{\bm{\theta}}_{k,n}\},~~\text{where}~ \widetilde{\bm{\theta}}_{k,i}=\bm{\theta}-\bm{\theta}_{k,i},\\
&\widetilde{\bar{\bm{\Theta}}}_{k}\overset{\triangle}{=}\text{col}\{\widetilde{\bar{\bm{\theta}}}_{k,1},\dots,\widetilde{\bar{\bm{\theta}}}_{k,n}\},~~\text{where}~ \widetilde{\bar{\bm{\theta}}}_{k,i}=\bm{\theta}-\bar{\bm{\theta}}_{k,i},\\
&\bm{P}_{k}\overset{\triangle}{=}\text{diag}\{P_{k,1},\dots,P_{k,n}\},~~\bar{\bm{P}}_{k}\overset{\triangle}{=}\text{diag}\{\bar{P}_{k,1},\dots,\bar{P}_{k,n}\},\\
&\bm{b}_{k}\overset{\triangle}{=}\text{diag}\{b_{k,1},\dots,b_{k,n}\},~~\bm{c}_{k}\overset{\triangle}{=}\bm{b}_{k}\otimes I_{m},~~\bm{\mathscr{A}}\overset{\triangle}{=}\mathcal{A}\otimes I_{m},
\end{aligned}
$$
where $\text{col}\{\cdots\}$ denotes a vector by stacking the specified vectors, $\text{diag}\{\cdots\}$ is used in a non-standard manner which means that $m\times 1$ column vectors are combined ``in a diagonal manner'' resulting in a $mn\times n$ matrix, and $\otimes$ is the Kronecker product. Then (\ref{model}) can be rewritten in the following compact form:
\begin{equation}\label{Model}
\bm{Y}_{k+1}=\bm{\Phi}_{k}^{\top}\bm{\Theta}+\bm{W}_{k+1},
\end{equation}
Similarly, for the distributed LS algorithm we have
\begin{equation}\label{7}
\begin{cases}
&\bar{\bm{\Theta}}_{k+1}=\bm{\Theta}_{k}+\bm{c}_{k}\bm{P}_{k}\bm{\Phi}_{k}(\bm{Y}_{k+1}-\bm{\Phi}_{k}^{\top}\bm{\Theta}_{k}),\\
&\bar{\bm{P}}_{k+1}=\bm{P}_{k}-\bm{c}_{k}\bm{P}_{k}\bm{\Phi}_{k}\bm{\Phi}_{k}^{\top}\bm{P}_{k},\\
&\bm{b}_{k}=(I_{n}+\bm{\Phi}_{k}^{\top}\bm{P}_{k}\bm{\Phi}_{k})^{-1},\\
&\bm{c}_{k}=\bm{b}_{k}\otimes I_{m},\\
&\text{vec}\{\bm{P}_{k+1}^{-1}\}=\mathscr{A}\text{vec}\{\bar{\bm{P}}_{k+1}^{-1}\},\\
&\bm{\Theta}_{k+1}=\bm{P}_{k+1}\mathscr{A}\bar{\bm{P}}_{k+1}^{-1}\bar{\bm{\Theta}}_{k+1},
\end{cases}
\end{equation}
where $\text{vec}\{\cdot\}$ denotes the operator that stacks the blocks of a block diagonal matrix on top of each other.

Since $\widetilde{\bm{\Theta}}_{k}=\bm{\Theta}-\bm{\Theta}_{k}$ and $\widetilde{\bar{\bm{\Theta}}}_{k}=\bm{\Theta}-\bar{\bm{\Theta}}_{k}$ by definition, substituting (\ref{Model}) into (\ref{7}), we can get
$$
\begin{aligned}
\widetilde{\bar{\bm{\Theta}}}_{k+1}
=&\bm{\Theta}-\bm{\Theta}_{k}-\bm{c}_{k}\bm{P}_{k}\bm{\Phi}_{k}(\bm{\Phi}_{k}^{\top}\bm{\Theta}+\bm{W}_{k+1}-\bm{\Phi}_{k}^{\top}\bm{\Theta}_{k})\\
=&(I_{mn}-\bm{c}_{k}\bm{P}_{k}\bm{\Phi}_{k}\bm{\Phi}_{k}^{\top})\widetilde{\bm{\Theta}}_{k}-\bm{c}_{k}\bm{P}_{k}\bm{\Phi}_{k}\bm{W}_{k+1}\\
=&\bar{\bm{P}}_{k+1}\bm{P}_{k}^{-1}\widetilde{\bm{\Theta}}_{k}-\bm{c}_{k}\bm{P}_{k}\bm{\Phi}_{k}\bm{W}_{k+1}.
\end{aligned}
$$
Note also that
$$
\begin{aligned}
&\bm{P}_{k+1}\mathscr{A}\bar{\bm{P}}_{k+1}^{-1}\bm{\Theta}\\
=&\text{col}\bigg\{P_{k+1,1}\sum_{j\in\mathcal{N}_{1}}a_{j1}\bar{P}_{k+1,j}^{-1}\bm{\theta},\dots,P_{k+1,n}\sum_{j\in\mathcal{N}_{n}}a_{jn}\bar{P}_{k+1,j}^{-1}\bm{\theta}\bigg\}\\
=&\bm{\Theta}.
\end{aligned}
$$
Then we have
\begin{align}\label{Theta}
\widetilde{\bm{\Theta}}_{k+1}
=&\bm{\Theta}-\bm{P}_{k+1}\mathscr{A}\bar{\bm{P}}_{k+1}^{-1}\bar{\bm{\Theta}}_{k+1}=\bm{P}_{k+1}\mathscr{A}\bar{\bm{P}}_{k+1}^{-1}\widetilde{\bar{\bm{\Theta}}}_{k+1}\nonumber\\
=&\bm{P}_{k+1}\mathscr{A}\bm{P}_{k}^{-1}\widetilde{\bm{\Theta}}_{k}\nonumber\\
&-\bm{P}_{k+1}\mathscr{A}\bar{\bm{P}}_{k+1}^{-1}\bm{c}_{k}\bm{P}_{k}\bm{\Phi}_{k}\bm{W}_{k+1}.
\end{align}

Before establishing the convergence of the distributed LS, we first present a critical theorem, which requires no excitation conditions on the regression process $\bm{\varphi}_{k,i}$.

\begin{thm}\label{thm1}
Let \emph{Assumptions \ref{asm1}} and \emph{\ref{asm2}} be satisfied, we have as $t\to\infty$,
\begin{enumerate}
\item
$
\sum\limits_{k=0}^{t}\widetilde{\bm{\Theta}}_{k}^{\top}\bm{\Phi}_{k}\bm{b}_{k}\bm{\Phi}_{k}^{\top}\widetilde{\bm{\Theta}}_{k}=O(\log (r_{t})),~~~\ \hbox{a.s.},
$

\item
$
\widetilde{\bm{\Theta}}_{t+1}^{\top}\bm{P}_{t+1}^{-1}\widetilde{\bm{\Theta}}_{t+1}=O(\log (r_{t})),~~~\ \hbox{a.s.},
$
\end{enumerate}
where
\begin{equation}\label{rtdef}
r_{t}=\lambda_{\max}\{\bm{P}_{0}^{-1}\}+\sum_{i=1}^{n}\sum_{k=0}^{t}\|\bm{\varphi}_{k,i}\|^{2}.
\end{equation}
\end{thm}

From this, we can obtain the upper bound of the accumulated regrets for the distributed LS-based adaptive predictor. For any $i\in\{1,\dots, n\}$, and at any time instant $k\geq 1$, the best prediction to the future observation $y_{k+1,i}$ is the conditional mathematical expectation $\mathbb{E}[y_{k+1,i}\vert \mathcal{F}_{k}]=\bm{\varphi}_{k,i}^{\top}\bm{\theta}$, since the noise is a martingale difference sequence with second moment.  Unfortunately, this optimal predictor is unavailable because $\bm{\theta}$ is unknown. A natural way is to construct an adaptive predictor $\widehat{y}_{k+1,i}$ by using the online distributed LS estimate $\bm{\theta}_{k,i}$, i.e., $\widehat{y}_{k+1,i}=\bm{\varphi}_{k,i}^{\top}\bm{\theta}_{k,i}$. The error between the best predictor and the adaptive predictor is referred to as the regret denoted by 
\begin{equation}
R_{k,i}=(\mathbb{E}[y_{k+1,i}\vert \mathcal{F}_{k}]-\widehat{y}_{k+1,i})^{2},
\end{equation} 
which may not be zero and even may not be small in sample paths due to the persistent disturbance of the unpredictable noises in the model. 

However, one may evaluate the averaged regrets defined as follows:
$$
\frac{1}{nt}\sum_{i=1}^{n}\sum_{k=0}^{t}R_{k,i}, 
$$
which we are going to show tends to zero as $t$ increases to infinity under essentially no excitation conditions on the regressors, see \emph{Theorem \ref{thm2}} below. This is a celebrated property that has been widely studied in distributed online learning and optimization problems \cite{Hos2016, Sha2018}, but under rather restrictive assumptions such as boundedness, stationarity or independence on the system signals.

\begin{thm}\label{thm2}
Let \emph{Assumptions \ref{asm1}} and \emph{\ref{asm2}} be satisfied. Then the sample paths of the accumulated regrets have the following bound as $t\to\infty$:
\begin{align}\label{Coro}
\sum_{i=1}^{n}\sum_{k=0}^{t}R_{k,i}=O(\log(r_{t})),~~~~\ \hbox{a.s.},
\end{align}
provided that $\bm{\Phi}_{t}^{\top}\bm{P}_{t}\bm{\Phi}_{t}=O(1), \ \hbox{a.s.}$
\end{thm}

\begin{rem}
We remark that the order $O(\log(r_{t}))$ for the accumulated regrets may be shown to be the best possible among all adaptive predictors in a certain sense, as is already known in the traditional single sensor case, see \cite{Lai1986A}. The precise constant in $O(\cdot)$ may also be determined if we have further conditions on the regressors, see \emph{Corollary 3.3} in \cite{Guo1995} in the single sensor case.
\end{rem}

From \emph{Theorem \ref{thm1}}, we can also obtain the strong consistency of the distributed LS to guarantee the generalization ability of learning.

\begin{thm}\label{thm3}
Let \emph{Assumptions \ref{asm1}}-\emph{\ref{asm3}} be satisfied, we have as $t\to\infty$,
\begin{equation}
\|\widetilde{\bm{\Theta}}_{t+1}\|^{2}=O\bigg(\frac{\log (r_{t})}{\lambda_{\min}^{n,t}}\bigg),~~\ \hbox{a.s.},
\end{equation}
where $r_{t}$ is defined by (\ref{rtdef}) and 
\begin{equation}\label{lambdamin}
\lambda_{\min}^{n,t}=\lambda_{\min}\bigg\{\sum_{j=1}^{n}P_{0,j}^{-1}+\sum_{j=1}^{n}\sum_{k=0}^{t-D_\mathcal{G}+1}\bm{\varphi}_{k,j}\bm{\varphi}_{k,j}^{\top}\bigg\}.
\end{equation}
\end{thm}

\begin{rem}
\emph{Theorem \ref{thm3}} shows that if
\begin{equation}\label{logrt}
\lim_{t\to\infty}\frac{\log (r_{t})}{\lambda_{\min}^{n,t}}=0,~~~~~\ \hbox{a.s.},
\end{equation}
then the distributed LS estimate $\bm{\Theta}_{t}$ will converge to the true unknown parameter. We may name (\ref{logrt}) as a cooperative excitation condition. In the traditional single sensor case (where $n=1$,  $D_{\mathcal{G}}=1$),  (\ref{logrt}) reduces to the well-known Lai-Wei excitation condition,  which is known to be the weakest possible data condition for the convergence of the classical LS estimates \cite{Lai1982}, and is much weaker than the well-known persistence of excitation (PE) condition usually used in the parameter estimation of finite-dimensional linear control
systems.

Moreover, it is easy to convince oneself that the cooperative excitation condition (\ref{logrt}) will make it possible for the distributed LS to consistently estimate the unknown parameter, even if any individual sensor cannot due to lack of suitable excitation, thanks to the cooperative nature of the excitation condition (\ref{logrt}). Finally, we remark that the verification of (\ref{logrt}) is straightforward in the ergodic case. For more general correlated non-stationary signals from control systems, the verification of (\ref{logrt}) may be conducted following a similar way as that for the traditional single sensor case (see,\cite{Chen1991}).

Furthermore, the convergence rate  established in \emph{Theorem \ref{thm3}} is essentially in terms of the increase of the number of observations rather than the number of iterations in computation. 
\end{rem}

\begin{rem}
Let us now compare the above distributed LS algorithm with centralized methods whereby, at each time instant $k$, all the $n$ sensors transmit their raw data $\{y_{k+1,i}, \bm{\varphi}_{k,i}\}$ to a fusion center for processing to obtain a centralized estimate. Although the centralized algorithm may have some advantages over the distributed algorithm in terms of communication complexity, it also has some drawbacks compared with the distributed case. Firstly, the distributed methods may have stronger structural robustness compared with the centralized ones. This is because the centralized algorithm will fail once the fusion center is broken down by outside attacks or some sensors lost the connection to the fusion center, while the distributed algorithm can still estimate the unknown parameters  even if the communications among some sensors are interrupted, as long as the network connectivity is maintained. Secondly, if the fusion center is far away from some sensors, the communications with the fusion center may not be feasible, and the transmission of observations and regression vectors may compromise the safety and privacy of the system even if the communication is possible. Hence, our distributed estimation problem is not a purely computational problem.
 \end{rem}

\section{Proofs of the main results}

\subsection{Proof of \emph{Theorem \ref{thm1}}}

To prove \emph{Theorem \ref{thm1}}, we need to establish several lemmas first. The first lemma below is a key inequality on convex combination of nonnegative definite matrices.

\begin{lem}\label{lem1}
For any adjacency matrix $\mathcal{A}=\{a_{ij}\}\in\mathbb{R}^{n\times n}$, denote $\mathscr{A}=\mathcal{A}\otimes I_{m}$, and for any nonnegative definite matrices $Q_{i}\in\mathbb{R}^{m\times m}, i=1,\dots,n$, denote $Q=\text{diag}\{Q_{1},\dots,Q_{n}\}$, and $Q^{'}=\text{diag}\{Q_{1}^{'},\dots,Q_{n}^{'}\}$, where $Q_{i}^{'}=\sum_{j=1}^{n}a_{ji}Q_{j}$. Then the following inequality holds: 
\begin{equation}
\mathscr{A}Q\mathscr{A}\leq Q^{'}.
\end{equation}
\end{lem}

\begin{IEEEproof}
By the definition of $\mathscr{A}$ and $Q$, we can get that
$$
\begin{aligned}
\mathscr{A}Q\mathscr{A}=\begin{pmatrix}
\sum\limits_{j=1}^{n}a_{1j}a_{j1}Q_{j} & \cdots & \sum\limits_{j=1}^{n}a_{1j}a_{jn}Q_{j}\\
\vdots & \ddots & \vdots\\
\sum\limits_{j=1}^{n}a_{nj}a_{j1}Q_{j} & \cdots & \sum\limits_{j=1}^{n}a_{nj}a_{jn}Q_{j}
\end{pmatrix}.
\end{aligned}
$$

In order to prove the inequality, we only need to prove that for any  unit column vector $x\in\mathbb{R}^{mn}$ with $\|x\|=1$, $x^{\top}\mathscr{A}Q\mathscr{A}x\leq x^{\top}Q^{'}x$ holds. Denote $x=\text{col}\{x_{1}, x_{2}, \dots, x_{n}\}$ with $x_{i}\in\mathbb{R}^{m}$, then by the Schwarz inequality and noticing that $Q_{j}\geq 0$, $\sum_{j=1}^{n}a_{ij}=1$, and $a_{ji}=a_{ij}, (i,j=1,\dots,n)$, we have
$$
\begin{aligned}
&x^{\top}\mathscr{A}Q\mathscr{A}x=\sum_{p=1}^{n}\sum_{q=1}^{n}\sum_{j=1}^{n}a_{pj}a_{jq}x_{p}^{\top}Q_{j}x_{q}\\
=&\sum_{p=1}^{n}\sum_{q=1}^{n}\sum_{j=1}^{n}\sqrt{a_{pj}a_{jq}}x_{p}^{\top}Q_{j}^{\frac{1}{2}}\cdot \sqrt{a_{pj}a_{jq}}Q_{j}^{\frac{1}{2}}x_{q}\\
\leq &\Bigg\{\sum_{p=1}^{n}\sum_{q=1}^{n}\sum_{j=1}^{n}a_{pj}a_{jq}x_{p}^{\top}Q_{j}x_{p}\Bigg\}^{\frac{1}{2}}\\
&\cdot\Bigg\{\sum_{p=1}^{n}\sum_{q=1}^{n}\sum_{j=1}^{n}a_{pj}a_{jq}x_{q}^{\top}Q_{j}x_{q}\Bigg\}^{\frac{1}{2}}\\
= &\Bigg\{\sum_{p=1}^{n}\sum_{j=1}^{n}a_{pj}x_{p}^{\top}Q_{j}x_{p}\Bigg\}^{\frac{1}{2}}\Bigg\{\sum_{q=1}^{n}\sum_{j=1}^{n}a_{jq}x_{q}^{\top}Q_{j}x_{q}\Bigg\}^{\frac{1}{2}}\\
=&\sum_{i=1}^{n}\sum_{j=1}^{n}a_{ji}x_{i}^{\top}Q_{j}x_{i}=x^{\top}Q^{'}x,
\end{aligned}
$$
which completes the proof.
\end{IEEEproof}

By \emph{Lemma \ref{lem1}}, we can obtain the following result:

\begin{lem}\label{lem2}
For any adjacency matrix $\mathcal{A}=\{a_{ij}\}\in\mathbb{R}^{n\times n}$, denote $\mathscr{A}=\mathcal{A}\otimes I_{m}$. Then for any $k\geq 1$,
\begin{equation}
\mathscr{A}\bar{\bm{P}}_{k+1}^{-1}\mathscr{A}\leq \bm{P}_{k+1}^{-1},
\end{equation}
and
\begin{equation}\label{APA}
\mathscr{A}\bm{P}_{k+1}\mathscr{A}\leq \bar{\bm{P}}_{k+1},
\end{equation}
holds, where $\bar{\bm{P}}_{k+1}$ and $\bm{P}_{k+1}$ are defined in (\ref{7}).
\end{lem}

\begin{IEEEproof}
By taking $Q_{i}=\bar{P}_{k+1,i}^{-1}\geq 0$ and noticing $P_{k+1,i}^{-1}=\sum_{j=1}^{n}a_{ji}\bar{P}_{k+1,j}^{-1}=Q_{i}^{'}$, we know  from \emph{Lemma 4.1} that
$$
\mathscr{A}\bar{\bm{P}}_{k+1}^{-1}\mathscr{A}\leq \bm{P}_{k+1}^{-1},
$$
holds. To prove (\ref{APA}), we first assume that $\mathscr{A}$ is invertible. Then by \emph{Lemma A.1} in Appendix A, it is easy to see that
$$
\mathscr{A}\bm{P}_{k+1}\mathscr{A}\leq \bar{\bm{P}}_{k+1}.
$$

Next, we consider the case where $\mathscr{A}$ is not invertible. Since the number of eigenvalues of the matrix $\mathscr{A}$ is finite,  then exists a constant $\varepsilon^{*}\in(0,1)$ such that the perturbed adjacency matrix $\mathscr{A}^{\varepsilon}=\mathscr{A}+\varepsilon I_{mn}=\{a_{ij}^{\varepsilon}\}$ will be invertible for any $0<\varepsilon<\varepsilon^{*}$. By the definition of $\mathscr{A}^{\varepsilon}$, we know that $\mathscr{A}^{\varepsilon}$ is symmetric and the sums of each columns and rows of the matrix $\mathscr{A}^{\varepsilon}$ are all $1+\varepsilon$. Then we define $(P_{k+1,i}^{\varepsilon})^{-1}=\sum_{j=1}^{n}a_{ji}^{\varepsilon}\bar{P}_{k+1,j}^{-1}$, and we can denote $\bm{P}_{k+1}^{\varepsilon}=\text{diag}\{P_{k+1,1}^{\varepsilon}, \dots, P_{k+1,n}^{\varepsilon}\}$ since $(P_{k+1,i}^{\varepsilon})^{-1}$ defined above is invertible. Similar to the proof of \emph{Lemma 4.1}, for any unit column vector $x\in\mathbb{R}^{mn}$, we have
$$
x^{\top}\mathscr{A}^{\varepsilon}\bar{\bm{P}}_{k+1}^{-1}\mathscr{A}^{\varepsilon}x\leq(1+\varepsilon)x^{\top}(\bm{P}_{k+1}^{\varepsilon})^{-1}x.
$$
Consequently, we have $\mathscr{A}^{\varepsilon}\bar{\bm{P}}_{k+1}^{-1}\mathscr{A}^{\varepsilon}\leq (1+\varepsilon)(\bm{P}_{k+1}^{\varepsilon})^{-1}$. Since $\mathscr{A}^{\varepsilon}$ is invertible, we know from \emph{Lemma A.1} in Appendix A that
$
\mathscr{A}^{\varepsilon}\bm{P}_{k+1}^{\varepsilon}\mathscr{A}^{\varepsilon}\leq (1+\varepsilon)\bar{\bm{P}}_{k+1}.
$
By taking $\varepsilon\to 0$ on both sides of the above equation, we can obtain that
$$
\lim_{\varepsilon\to 0}\mathscr{A}^{\varepsilon}\bm{P}_{k+1}^{\varepsilon}\mathscr{A}^{\varepsilon}=\mathscr{A}\bm{P}_{k+1}\mathscr{A}\leq\lim_{\varepsilon\to 0}(1+\varepsilon)\bar{\bm{P}}_{k+1}=\bar{\bm{P}}_{k+1}.
$$
This completes the proof.
\end{IEEEproof}

To accomplish the proof of \emph{Theorem \ref{thm1}}, we also need the following inequality:

\begin{lem}\label{lem3}
For any adjacency matrix $\mathcal{A}=\{a_{ij}\}\in\mathbb{R}^{n\times n}$, and for any $k\geq 1$, 
\begin{equation}
\vert\bar{\bm{P}}_{k+1}^{-1}\vert\leq\vert\bm{P}_{k+1}^{-1}\vert,
\end{equation} 
holds, where $\bar{\bm{P}}_{k+1}$ and $\bm{P}_{k+1}$ are defined in (\ref{7}).
\end{lem}

\begin{IEEEproof}
By Ky Fan convex theorem \cite{Ky1950} and noticing the definitions of $\bar{\bm{P}}_{k+1}, \bm{P}_{k+1}$, and $\mathcal{A}=\{a_{ij}\}$, we can see that
$$
\begin{aligned}
\vert\bm{P}_{k+1}^{-1}\vert=&\prod_{i=1}^{n}\bigg\vert\sum_{j=1}^{n}a_{ji}\bar{P}_{k+1,j}^{-1}\bigg\vert\\
\geq & \prod_{i=1}^{n}\vert\bar{P}_{k+1,1}^{-1}\vert^{a_{1i}}\vert\bar{P}_{k+1,2}^{-1}\vert^{a_{2i}}\cdots\vert\bar{P}_{k+1,n}^{-1}\vert^{a_{ni}}\\
=&\vert\bar{P}_{k+1,1}^{-1}\vert^{\sum\limits_{i=1}^{n}a_{1i}}\vert\bar{P}_{k+1,2}^{-1}\vert^{\sum\limits_{i=1}^{n}a_{2i}}\cdots\vert\bar{P}_{k+1,n}^{-1}\vert^{\sum\limits_{i=1}^{n}a_{ni}}\\
=&\vert\bar{P}_{k+1,1}^{-1}\vert\cdot\vert\bar{P}_{k+1,2}^{-1}\vert\cdots\vert\bar{P}_{k+1,n}^{-1}\vert=\vert\bar{\bm{P}}_{k+1}^{-1}\vert,
\end{aligned}
$$
which completes the proof.
\end{IEEEproof}

To prove \emph{Theorem \ref{thm1}}, we also need the following critical lemma:

\begin{lem}\label{lem4}
Let \emph{Assumptions \ref{asm1}} and \emph{\ref{asm2}} be satisfied. Then the distributed LS defined by (\ref{Model}) and (\ref{7}) satisfies the following relationship as $t\to\infty$:
\begin{align}\label{lemma4.4}
&\widetilde{\bm{\Theta}}_{t+1}^{\top}\bm{P}_{t+1}^{-1}\widetilde{\bm{\Theta}}_{t+1}+[1+o(1)]\sum_{k=0}^{t}\widetilde{\bm{\Theta}}_{k}^{\top}\bm{\Phi}_{k}\bm{b}_{k}\bm{\Phi}_{k}^{\top}\widetilde{\bm{\Theta}}_{k}\nonumber\\
&+[1+o(1)]\sum_{k=0}^{t}\widetilde{\bm{\Theta}}_{k}^{\top}\bm{P}_{k}^{-1}\bm{\Delta}_{k+1}\bm{P}_{k}^{-1}\widetilde{\bm{\Theta}}_{k}\nonumber\\
\leq &\sigma_{w}\log(\vert\bm{P}_{t+1}^{-1}\vert)+o(\log(\vert\bm{P}_{t+1}^{-1}\vert))+O(1),~~\ \hbox{a.s.},
\end{align}
where $\bm{\Delta}_{k+1}\overset{\triangle}{=}\bar{\bm{P}}_{k+1}-\mathscr{A}\bm{P}_{k+1}\mathscr{A}\geq 0$ by \emph{Lemma \ref{lem2}}, and $\sigma_{w}=\sum_{i=1}^{n}\sigma_{i}^{2}, \sigma_{i}^{2}=\sup_{k\geq 0}\mathbb{E}[w_{k+1,i}^{2}\vert\mathcal{F}_{k}]$.
\end{lem}

\begin{IEEEproof}
Since $\bm{b}_{k}=(I_{n}+\bm{\Phi}_{k}^{\top}\bm{P}_{k}\bm{\Phi}_{k})^{-1}$ and $\bm{c}_{k}=\bm{b}_{k}\otimes I_{m}$, then by (\ref{Theta}), we know that 
$$
\widetilde{\bm{\Theta}}_{k+1}=\bm{P}_{k+1}\mathscr{A}\bm{P}_{k}^{-1}\widetilde{\bm{\Theta}}_{k}-\bm{P}_{k+1}\mathscr{A}\bar{\bm{P}}_{k+1}^{-1}\bm{c}_{k}\bm{P}_{k}\bm{\Phi}_{k}\bm{W}_{k+1}. 
$$
Hence, we have the following expansion for the stochastic Lyapunov function $V_{k}=\widetilde{\bm{\Theta}}_{k}^{\top}\bm{P}_{k}^{-1}\widetilde{\bm{\Theta}}_{k}$:
\begin{align}\label{Lya1}
V_{k+1}=&\widetilde{\bm{\Theta}}_{k+1}^{\top}\bm{P}_{k+1}^{-1}\widetilde{\bm{\Theta}}_{k+1}\nonumber\\
=&(\widetilde{\bm{\Theta}}_{k}^{\top}\bm{P}_{k}^{-1}\mathscr{A}\bm{P}_{k+1}-\bm{W}_{k+1}^{\top}\bm{\Phi}_{k}^{\top}\bm{P}_{k}\bm{c}_{k}\bar{\bm{P}}_{k+1}^{-1}\mathscr{A}\bm{P}_{k+1})\nonumber\\
&\cdot(\mathscr{A}\bm{P}_{k}^{-1}\widetilde{\bm{\Theta}}_{k}-\mathscr{A}\bar{\bm{P}}_{k+1}^{-1}\bm{c}_{k}\bm{P}_{k}\bm{\Phi}_{k}\bm{W}_{k+1})\nonumber\\
=&\widetilde{\bm{\Theta}}_{k}^{\top}\bm{P}_{k}^{-1}\mathscr{A}\bm{P}_{k+1}\mathscr{A}\bm{P}_{k}^{-1}\widetilde{\bm{\Theta}}_{k}\nonumber\\
&-2\widetilde{\bm{\Theta}}_{k}^{\top}\bm{P}_{k}^{-1}\mathscr{A}\bm{P}_{k+1}\mathscr{A}\bar{\bm{P}}_{k+1}^{-1}\bm{c}_{k}\bm{P}_{k}\bm{\Phi}_{k}\bm{W}_{k+1}\nonumber\\
&+\bm{W}_{k+1}^{\top}\bm{\Phi}_{k}^{\top}\bm{P}_{k}\bm{c}_{k}\bar{\bm{P}}_{k+1}^{-1}\mathscr{A}\bm{P}_{k+1}\mathscr{A}\bar{\bm{P}}_{k+1}^{-1}\nonumber\\
&\cdot\bm{c}_{k}\bm{P}_{k}\bm{\Phi}_{k}\bm{W}_{k+1}.
\end{align}

Now, we proceed to estimate the right-hand-side (RHS) of (\ref{Lya1}) term by term. Firstly, we know that
\begin{align}\label{term1}
&\widetilde{\bm{\Theta}}_{k}^{\top}\bm{P}_{k}^{-1}\mathscr{A}\bm{P}_{k+1}\mathscr{A}\bm{P}_{k}^{-1}\widetilde{\bm{\Theta}}_{k}\nonumber\\
=&\widetilde{\bm{\Theta}}_{k}^{\top}\bm{P}_{k}^{-1}\bar{\bm{P}}_{k+1}\bm{P}_{k}^{-1}\widetilde{\bm{\Theta}}_{k}-\widetilde{\bm{\Theta}}_{k}^{\top}\bm{P}_{k}^{-1}\bm{\Delta}_{k+1}\bm{P}_{k}^{-1}\widetilde{\bm{\Theta}}_{k}\nonumber\\
=&\widetilde{\bm{\Theta}}_{k}^{\top}\bm{P}_{k}^{-1}(\bm{P}_{k}-\bm{P}_{k}\bm{\Phi}_{k}\bm{b}_{k}\bm{\Phi}_{k}^{\top}\bm{P}_{k})\bm{P}_{k}^{-1}\widetilde{\bm{\Theta}}_{k}\nonumber\\
&-\widetilde{\bm{\Theta}}_{k}^{\top}\bm{P}_{k}^{-1}\bm{\Delta}_{k+1}\bm{P}_{k}^{-1}\widetilde{\bm{\Theta}}_{k}\nonumber\\
=&\widetilde{\bm{\Theta}}_{k}^{\top}\bm{P}_{k}^{-1}\widetilde{\bm{\Theta}}_{k}-\widetilde{\bm{\Theta}}_{k}^{\top}\bm{\Phi}_{k}\bm{b}_{k}\bm{\Phi}_{k}^{\top}\widetilde{\bm{\Theta}}_{k}-\widetilde{\bm{\Theta}}_{k}^{\top}\bm{P}_{k}^{-1}\bm{\Delta}_{k+1}\bm{P}_{k}^{-1}\widetilde{\bm{\Theta}}_{k}\nonumber\\
=&V_{k}-\widetilde{\bm{\Theta}}_{k}^{\top}\bm{\Phi}_{k}\bm{b}_{k}\bm{\Phi}_{k}^{\top}\widetilde{\bm{\Theta}}_{k}-\widetilde{\bm{\Theta}}_{k}^{\top}\bm{P}_{k}^{-1}\bm{\Delta}_{k+1}\bm{P}_{k}^{-1}\widetilde{\bm{\Theta}}_{k}.
\end{align}

Moreover, by the (block) diagonal property of $\bm{b}_{k}, \bm{c}_{k}, \bm{P}_{k}$ and $\bm{\Phi}_{k}$, we have $\bm{c}_{k}\bm{P}_{k}=\bm{P}_{k}\bm{c}_{k}, \bm{\Phi}_{k}^{\top}\bm{c}_{k}=\bm{b}_{k}\bm{\Phi}_{k}^{\top}$, and $\bm{c}_{k}\bm{\Phi}_{k}=\bm{\Phi}_{k}\bm{b}_{k}$.
By the matrix inversion lemma \cite{Guo},  we have 
$$
\bar{\bm{P}}_{k+1}^{-1}=\bm{P}_{k}^{-1}+\bm{\Phi}_{k}\bm{\Phi}_{k}^{\top}. 
$$
Thus, we can estimate the second term on the RHS of (\ref{Lya1}) as follows:
\begin{align}\label{term2}
&\widetilde{\bm{\Theta}}_{k}^{\top}\bm{P}_{k}^{-1}\mathscr{A}\bm{P}_{k+1}\mathscr{A}\bar{\bm{P}}_{k+1}^{-1}\bm{c}_{k}\bm{P}_{k}\bm{\Phi}_{k}\bm{W}_{k+1}\nonumber\\
=&\widetilde{\bm{\Theta}}_{k}^{\top}\bm{P}_{k}^{-1}\mathscr{A}\bm{P}_{k+1}\mathscr{A}\bm{c}_{k}\bm{\Phi}_{k}\bm{W}_{k+1}\nonumber\\
&+\widetilde{\bm{\Theta}}_{k}^{\top}\bm{P}_{k}^{-1}\mathscr{A}\bm{P}_{k+1}\mathscr{A}\bm{\Phi}_{k}\bm{\Phi}_{k}^{\top}\bm{c}_{k}\bm{P}_{k}\bm{\Phi}_{k}\bm{W}_{k+1}\nonumber\\
=&\widetilde{\bm{\Theta}}_{k}^{\top}\bm{P}_{k}^{-1}\mathscr{A}\bm{P}_{k+1}\mathscr{A}\bm{c}_{k}\bm{\Phi}_{k}\bm{W}_{k+1}\nonumber\\
&+\widetilde{\bm{\Theta}}_{k}^{\top}\bm{P}_{k}^{-1}\mathscr{A}\bm{P}_{k+1}\mathscr{A}\bm{\Phi}_{k}\bm{W}_{k+1}\nonumber\\
&-\widetilde{\bm{\Theta}}_{k}^{\top}\bm{P}_{k}^{-1}\mathscr{A}\bm{P}_{k+1}\mathscr{A}\bm{\Phi}_{k}\bm{b}_{k}\bm{W}_{k+1}\nonumber\\
=&\widetilde{\bm{\Theta}}_{k}^{\top}\bm{P}_{k}^{-1}\mathscr{A}\bm{P}_{k+1}\mathscr{A}\bm{\Phi}_{k}\bm{W}_{k+1}\nonumber\\
=&\widetilde{\bm{\Theta}}_{k}^{\top}\bm{P}_{k}^{-1}\bar{\bm{P}}_{k+1}\bm{\Phi}_{k}\bm{W}_{k+1}-\widetilde{\bm{\Theta}}_{k}^{\top}\bm{P}_{k}^{-1}\bm{\Delta}_{k+1}\bm{\Phi}_{k}\bm{W}_{k+1}.
\end{align}

As for the last term on the RHS of (\ref{Lya1}), by $\mathscr{A}\bm{P}_{k+1}\mathscr{A}\leq \bar{\bm{P}}_{k+1}$, we can estimate it as follows:
\begin{align}\label{term3}
&\bm{W}_{k+1}^{\top}\bm{\Phi}_{k}^{\top}\bm{P}_{k}\bm{c}_{k}\bar{\bm{P}}_{k+1}^{-1}\mathscr{A}\bm{P}_{k+1}\mathscr{A}\bar{\bm{P}}_{k+1}^{-1}\bm{c}_{k}\bm{P}_{k}\bm{\Phi}_{k}\bm{W}_{k+1}\nonumber\\
\leq &\bm{W}_{k+1}^{\top}\bm{\Phi}_{k}^{\top}\bm{P}_{k}\bm{c}_{k}(\bm{P}_{k}^{-1}+\bm{\Phi}_{k}\bm{\Phi}_{k}^{\top})\bm{c}_{k}\bm{P}_{k}\bm{\Phi}_{k}\bm{W}_{k+1}\nonumber\\
=&\bm{W}_{k+1}^{\top}\bm{\Phi}_{k}^{\top}\bm{P}_{k}\bm{c}_{k}^{2}\bm{\Phi}_{k}\bm{W}_{k+1}\nonumber\\
&+\bm{W}_{k+1}^{\top}\bm{\Phi}_{k}^{\top}\bm{P}_{k}\bm{c}_{k}\bm{\Phi}_{k}\bm{\Phi}_{k}^{\top}\bm{c}_{k}\bm{P}_{k}\bm{\Phi}_{k}\bm{W}_{k+1}\nonumber\\
=&\bm{W}_{k+1}^{\top}\bm{b}_{k}^{2}\bm{\Phi}_{k}^{\top}\bm{P}_{k}\bm{\Phi}_{k}\bm{W}_{k+1}\nonumber\\
&+\bm{W}_{k+1}^{\top}(I_{n}+\bm{\Phi}_{k}^{\top}\bm{P}_{k}\bm{\Phi}_{k})\bm{b}_{k}^{2}\bm{\Phi}_{k}^{\top}\bm{P}_{k}\bm{\Phi}_{k}\bm{W}_{k+1}\nonumber\\
&-\bm{W}_{k+1}^{\top}\bm{b}_{k}^{2}\bm{\Phi}_{k}^{\top}\bm{P}_{k}\bm{\Phi}_{k}\bm{W}_{k+1}\nonumber\\
=&\bm{W}_{k+1}^{\top}\bm{b}_{k}\bm{\Phi}_{k}^{\top}\bm{P}_{k}\bm{\Phi}_{k}\bm{W}_{k+1}.
\end{align}
By (\ref{Lya1})-(\ref{term3}), we have 
\begin{align}\label{Lya}
V_{k+1}\leq&V_{k}-\widetilde{\bm{\Theta}}_{k}^{\top}\bm{\Phi}_{k}\bm{b}_{k}\bm{\Phi}_{k}^{\top}\widetilde{\bm{\Theta}}_{k}-\widetilde{\bm{\Theta}}_{k}^{\top}\bm{P}_{k}^{-1}\bm{\Delta}_{k+1}\bm{P}_{k}^{-1}\widetilde{\bm{\Theta}}_{k}\nonumber\\
&-2\widetilde{\bm{\Theta}}_{k}^{\top}\bm{P}_{k}^{-1}\bar{\bm{P}}_{k+1}\bm{\Phi}_{k}\bm{W}_{k+1}\nonumber\\
&+2\widetilde{\bm{\Theta}}_{k}^{\top}\bm{P}_{k}^{-1}\bm{\Delta}_{k+1}\bm{\Phi}_{k}\bm{W}_{k+1}\nonumber\\
&+\bm{W}_{k+1}^{\top}\bm{b}_{k}\bm{\Phi}_{k}^{\top}\bm{P}_{k}\bm{\Phi}_{k}\bm{W}_{k+1}.
\end{align}
Summing from $k=0$ to $t$ yields
\begin{align}\label{sum}
&V_{t+1}+\sum_{k=0}^{t}\widetilde{\bm{\Theta}}_{k}^{\top}\bm{\Phi}_{k}\bm{b}_{k}\bm{\Phi}_{k}^{\top}\widetilde{\bm{\Theta}}_{k}+\sum_{k=0}^{t}\widetilde{\bm{\Theta}}_{k}^{\top}\bm{P}_{k}^{-1}\bm{\Delta}_{k+1}\bm{P}_{k}^{-1}\widetilde{\bm{\Theta}}_{k}\nonumber\\
\leq & V_{0}-2\sum_{k=0}^{t}\widetilde{\bm{\Theta}}_{k}^{\top}\bm{P}_{k}^{-1}\bar{\bm{P}}_{k+1}\bm{\Phi}_{k}\bm{W}_{k+1}\nonumber\\
&-2\sum_{k=0}^{t}\widetilde{\bm{\Theta}}_{k}^{\top}\bm{P}_{k}^{-1}(-\bm{\Delta}_{k+1})\bm{\Phi}_{k}\bm{W}_{k+1}\nonumber\\
&+\sum_{k=0}^{t}\bm{W}_{k+1}^{\top}\bm{b}_{k}\bm{\Phi}_{k}^{\top}\bm{P}_{k}\bm{\Phi}_{k}\bm{W}_{k+1}.
\end{align}
Next, we estimate the last three terms on the RHS of (\ref{sum}) separately. By \emph{Assumptions \ref{asm1}} and \emph{\ref{asm2}}, and $\widetilde{\bm{\Theta}}_{k}^{\top}\bm{P}_{k}^{-1}\bar{\bm{P}}_{k+1}\bm{\Phi}_{k}\in\mathcal{F}_{k}, \widetilde{\bm{\Theta}}_{k}^{\top}\bm{P}_{k}^{-1}(-\bm{\Delta}_{k+1})\bm{\Phi}_{k}\in\mathcal{F}_{k}$, we can use the martingale estimation theorem (\emph{Theorem 2.8} in \cite{Chen1991}) to get the following estimation for any $\delta>0$:
\begin{align}\label{estimate1}
&\sum_{k=0}^{t}\widetilde{\bm{\Theta}}_{k}^{\top}\bm{P}_{k}^{-1}\bar{\bm{P}}_{k+1}\bm{\Phi}_{k}\bm{W}_{k+1}\nonumber\\
=&O\bigg(\bigg\{\sum_{k=0}^{t}\|\widetilde{\bm{\Theta}}_{k}^{\top}\bm{P}_{k}^{-1}\bar{\bm{P}}_{k+1}\bm{\Phi}_{k}\|^{2}\bigg\}^{\frac{1}{2}+\delta}\bigg),~~\ \hbox{a.s.},
\end{align}
and
\begin{align}\label{estimate2}
&\sum_{k=0}^{t}\widetilde{\bm{\Theta}}_{k}^{\top}\bm{P}_{k}^{-1}(-\bm{\Delta}_{k+1})\bm{\Phi}_{k}\bm{W}_{k+1}\nonumber\\
=&O\bigg(\bigg\{\sum_{k=0}^{t}\|\widetilde{\bm{\Theta}}_{k}^{\top}\bm{P}_{k}^{-1}\bm{\Delta}_{k+1}\bm{\Phi}_{k}\|^{2}\bigg\}^{\frac{1}{2}+\delta}\bigg),~~\ \hbox{a.s.}
\end{align}

To further analyze (\ref{estimate1}) and (\ref{estimate2}), we note from the definitions of $\bar{\bm{P}}_{k+1}$ and $\bm{b}_{k}$ that 
$$
\begin{aligned}
&\bm{P}_{k}^{-1}\bar{\bm{P}}_{k+1}\bm{\Phi}_{k}=\bm{\Phi}_{k}-\bm{c}_{k}\bm{\Phi}_{k}\bm{\Phi}_{k}^{\top}\bm{P}_{k}\bm{\Phi}_{k}\\
=&\bm{\Phi}_{k}-\bm{c}_{k}\bm{\Phi}_{k}(I_{n}+\bm{\Phi}_{k}^{\top}\bm{P}_{k}\bm{\Phi}_{k})+\bm{c}_{k}\bm{\Phi}_{k}=\bm{\Phi}_{k}\bm{b}_{k}.
\end{aligned}
$$
Hence, it is easy to see that
$$
\begin{aligned}
\|\widetilde{\bm{\Theta}}_{k}^{\top}\bm{P}_{k}^{-1}\bar{\bm{P}}_{k+1}\bm{\Phi}_{k}\|^{2}=&\widetilde{\bm{\Theta}}_{k}^{\top}\bm{P}_{k}^{-1}\bar{\bm{P}}_{k+1}\bm{\Phi}_{k}\bm{\Phi}_{k}^{\top}\bar{\bm{P}}_{k+1}\bm{P}_{k}^{-1}\widetilde{\bm{\Theta}}_{k}\\
=&\widetilde{\bm{\Theta}}_{k}^{\top}\bm{\Phi}_{k}\bm{b}_{k}^{2}\bm{\Phi}_{k}^{\top}\widetilde{\bm{\Theta}}_{k}\leq \widetilde{\bm{\Theta}}_{k}^{\top}\bm{\Phi}_{k}\bm{b}_{k}\bm{\Phi}_{k}^{\top}\widetilde{\bm{\Theta}}_{k},
\end{aligned}
$$
By taking $0<\delta<\frac{1}{2}$, we have from (\ref{estimate1}) that
\begin{align}\label{estimate1new}
&\sum_{k=0}^{t}\widetilde{\bm{\Theta}}_{k}^{\top}\bm{P}_{k}^{-1}\bar{\bm{P}}_{k+1}\bm{\Phi}_{k}\bm{W}_{k+1}\nonumber\\
=&O(1)+o\bigg(\bigg\{\sum_{k=0}^{t}\|\widetilde{\bm{\Theta}}_{k}^{\top}\bm{P}_{k}^{-1}\bar{\bm{P}}_{k+1}\bm{\Phi}_{k}\|^{2}\bigg\}\bigg)\nonumber\\
=&O(1)+o\bigg(\bigg\{\sum_{k=0}^{t}\widetilde{\bm{\Theta}}_{k}^{\top}\bm{\Phi}_{k}\bm{b}_{k}\bm{\Phi}_{k}^{\top}\widetilde{\bm{\Theta}}_{k}\bigg\}\bigg),~~~\ \hbox{a.s.}
\end{align}

Moreover, since $\bm{\Delta}_{k+1}=\bar{\bm{P}}_{k+1}-\mathscr{A}\bm{P}_{k+1}\mathscr{A}\leq \bar{\bm{P}}_{k+1}$, then
$$
\begin{aligned}
&\bm{\Delta}_{k+1}^{\frac{1}{2}}\bm{\Phi}_{k}\bm{\Phi}_{k}^{\top}\bm{\Delta}_{k+1}^{\frac{1}{2}}\leq\lambda_{\max}\{\bm{\Phi}_{k}^{\top}\bm{\Delta}_{k+1}\bm{\Phi}_{k}\}\cdot I_{mn}\\\leq &\lambda_{\max}\{\bm{\Phi}_{k}^{\top}\bar{\bm{P}}_{k+1}\bm{\Phi}_{k}\}\cdot I_{mn}\\
=&\lambda_{\max}\{\bm{\Phi}_{k}^{\top}(\bm{P}_{k}-\bm{c}_{k}\bm{P}_{k}\bm{\Phi}_{k}\bm{\Phi}_{k}^{\top}\bm{P}_{k})\bm{\Phi}_{k}\}\cdot I_{mn}\\
=&\lambda_{\max}\{\bm{b}_{k}\bm{\Phi}_{k}^{\top}\bm{P}_{k}\bm{\Phi}_{k}\}\cdot I_{mn}<I_{mn}.
\end{aligned}
$$
Hence, we have $\bm{\Delta}_{k+1}\bm{\Phi}_{k}\bm{\Phi}_{k}^{\top}\bm{\Delta}_{k+1}\leq \bm{\Delta}_{k+1}$, and so we have
\begin{align}\label{es2}
\|\widetilde{\bm{\Theta}}_{k}^{\top}\bm{P}_{k}^{-1}\bm{\Delta}_{k+1}\bm{\Phi}_{k}\|^{2}=&\widetilde{\bm{\Theta}}_{k}^{\top}\bm{P}_{k}^{-1}\bm{\Delta}_{k+1}\bm{\Phi}_{k}\bm{\Phi}_{k}^{\top}\bm{\Delta}_{k+1}\bm{P}_{k}^{-1}\widetilde{\bm{\Theta}}_{k}\nonumber\\
\leq & \widetilde{\bm{\Theta}}_{k}^{\top}\bm{P}_{k}^{-1}\bm{\Delta}_{k+1}\bm{P}_{k}^{-1}\widetilde{\bm{\Theta}}_{k}.
\end{align}
By taking $0<\delta<\frac{1}{2}$, we know from (\ref{estimate2}) that
\begin{align}\label{estimate2new}
&\sum_{k=0}^{t}\widetilde{\bm{\Theta}}_{k}^{\top}\bm{P}_{k}^{-1}(-\bm{\Delta}_{k+1})\bm{\Phi}_{k}\bm{W}_{k+1}\nonumber\\
=&O(1)+o\bigg(\bigg\{\sum_{k=0}^{t}\|\widetilde{\bm{\Theta}}_{k}^{\top}\bm{P}_{k}^{-1}\bm{\Delta}_{k+1}\bm{\Phi}_{k}\|^{2}\bigg\}\bigg)\nonumber\\
=&O(1)+o\bigg(\bigg\{\sum_{k=0}^{t}\widetilde{\bm{\Theta}}_{k}^{\top}\bm{P}_{k}^{-1}\bm{\Delta}_{k+1}\bm{P}_{k}^{-1}\widetilde{\bm{\Theta}}_{k}\bigg\}\bigg),~~~\ \hbox{a.s.}
\end{align}

We now proceed to estimate the last term in (\ref{sum}). Firstly, we know that
\begin{align}\label{last}
&\bm{W}_{k+1}^{\top}\bm{b}_{k}\bm{\Phi}_{k}^{\top}\bm{P}_{k}\bm{\Phi}_{k}\bm{W}_{k+1}\leq\|\bm{b}_{k}\bm{\Phi}_{k}^{\top}\bm{P}_{k}\bm{\Phi}_{k}\|\cdot\|\bm{W}_{k+1}\|^{2}\nonumber\\
=&\lambda_{\max}\{\bm{b}_{k}\bm{\Phi}_{k}^{\top}\bm{P}_{k}\bm{\Phi}_{k}\}\cdot\bigg\{\sum_{i=1}^{n}w_{k+1,i}^{2}\bigg\}.
\end{align}
Following a similar proof idea as in the traditional single sensor case (\cite{Lai1982}, see also \cite{Chen1991}), from $\bar{\bm{P}}_{k+1}=\bm{P}_{k}-\bm{c}_{k}\bm{P}_{k}\bm{\Phi}_{k}\bm{\Phi}_{k}^{\top}\bm{P}_{k}$, we have $\bm{P}_{k}^{-1}=\bar{\bm{P}}_{k+1}^{-1}(I_{mn}-\bm{c}_{k}\bm{P}_{k}\bm{\Phi}_{k}\bm{\Phi}_{k}^{\top})$. By taking determinants on both sides of the above identity, and noticing $0\leq\bm{b}_{k}\bm{\Phi}_{k}^{\top}\bm{P}_{k}\bm{\Phi}_{k}\leq I_{n}$, we have
$$
\begin{aligned}
\vert\bm{P}_{k}^{-1}\vert
=&\vert\bar{\bm{P}}_{k+1}^{-1}\vert\cdot\vert I_{n}-\bm{b}_{k}\bm{\Phi}_{k}^{\top}\bm{P}_{k}\bm{\Phi}_{k}\vert\\
\leq &\vert\bar{\bm{P}}_{k+1}^{-1}\vert\cdot(1-\lambda_{\max}\{\bm{b}_{k}\bm{\Phi}_{k}^{\top}\bm{P}_{k}\bm{\Phi}_{k}\}).
\end{aligned}
$$

Moreover, we know from \emph{Lemma \ref{lem3}} that
$$
\begin{aligned}
\lambda_{\max}\{\bm{b}_{k}\bm{\Phi}_{k}^{\top}\bm{P}_{k}\bm{\Phi}_{k}\}\leq \frac{\vert\bar{\bm{P}}_{k+1}^{-1}\vert-\vert\bm{P}_{k}^{-1}\vert}{\vert\bar{\bm{P}}_{k+1}^{-1}\vert}\leq \frac{\vert\bm{P}_{k+1}^{-1}\vert-\vert\bm{P}_{k}^{-1}\vert}{\vert\bm{P}_{k+1}^{-1}\vert}.
\end{aligned}
$$
Therefore
\begin{align}\label{log}
&\sum_{k=0}^{t}\lambda_{\max}\{\bm{b}_{k}\bm{\Phi}_{k}^{\top}\bm{P}_{k}\bm{\Phi}_{k}\}\leq\sum_{k=0}^{t}\frac{\vert\bm{P}_{k+1}^{-1}\vert-\vert\bm{P}_{k}^{-1}\vert}{\vert\bm{P}_{k+1}^{-1}\vert}\nonumber\\
\leq &\sum_{k=0}^{t}\int_{\vert\bm{P}_{k}^{-1}\vert}^{\vert\bm{P}_{k+1}^{-1}\vert}\frac{dx}{x}=\log(\vert\bm{P}_{t+1}^{-1}\vert)-\log(\vert\bm{P}_{0}^{-1}\vert).
\end{align}

Consequently, by using the martingale estimation theorem (\emph{Theorem 2.8} in \cite{Chen1991}), we have for any $\forall\eta>0$,
\begin{align}\label{lambdamax}
&\sum_{k=0}^{t}\lambda_{\max}\{\bm{b}_{k}\bm{\Phi}_{k}^{\top}\bm{P}_{k}\bm{\Phi}_{k}\}\bigg\{\sum_{i=1}^{n}w_{k+1,i}^{2}-\mathbb{E}\bigg[\sum_{i=1}^{n}w_{k+1,i}^{2}\vert\mathcal{F}_{k}\bigg]\bigg\}\nonumber\\
=&O\bigg(S_{t}\bigg(\frac{\beta}{2}\bigg)\bigg\{\log\bigg(S_{t}\bigg(\frac{\beta}{2}\bigg)+e\bigg)\bigg\}^{\frac{2}{\beta}+\eta}\bigg),~~\ \hbox{a.s.},
\end{align}
where
$$
S_{t}\bigg(\frac{\beta}{2}\bigg)\overset{\triangle}{=}\bigg[\sum_{k=0}^{t}(\lambda_{\max}\{\bm{b}_{k}\bm{\Phi}_{k}^{\top}\bm{P}_{k}\bm{\Phi}_{k}\})^{\frac{\beta}{2}}\bigg]^{\frac{2}{\beta}}.
$$
Since $\bm{b}_{k}\bm{\Phi}_{k}^{\top}\bm{P}_{k}\bm{\Phi}_{k}\leq I_{n}$ and $\frac{\beta}{2}>1$, we have from (\ref{log}) that 
$$
S_{t}\bigg(\frac{\beta}{2}\bigg)=O(1)+O((\log\vert\bm{P}_{t+1}^{-1}\vert)^{\frac{2}{\beta}}).
$$ 
From this, we can get from (\ref{last})-(\ref{lambdamax}) that
$$
\begin{aligned}
&\sum_{k=0}^{t}\bm{W}_{k+1}^{\top}\bm{b}_{k}\bm{\Phi}_{k}^{\top}\bm{P}_{k}\bm{\Phi}_{k}\bm{W}_{k+1}\\
\leq&\sum_{i=1}^{n}\sigma_{i}^{2}\sum_{k=0}^{t}\lambda_{\max}\{\bm{b}_{k}\bm{\Phi}_{k}^{\top}\bm{P}_{k}\bm{\Phi}_{k}\}+o(\log\vert\bm{P}_{t+1}^{-1}\vert)+O(1)\\
\leq &\sigma_{w}\log\vert\bm{P}_{t+1}^{-1}\vert+o(\log\vert\bm{P}_{t+1}^{-1}\vert)+O(1).
\end{aligned}
$$
Finally, substituting this into (\ref{sum}), we know that the desired result (\ref{lemma4.4}) is true. This completes the proof.
\end{IEEEproof}

\textbf{\emph{Proof of Theorem 3.1}}:

By the definitions of $\bar{P}_{t,i}^{-1}$ and $P_{t,i}^{-1}$, it is easy to know that for any $t\geq 0$, $$
P_{t+1,i}^{-1}=\sum_{j=1}^{n}a_{ji}\bar{P}_{t+1,j}^{-1}=\sum_{j=1}^{n}a_{ji}(P_{t,j}^{-1}+\bm{\varphi}_{t,j}\bm{\varphi}_{t,j}^{\top}). 
$$
Consequently, we have
\begin{align}\label{Pt}
&\max_{1\leq i\leq n}\lambda_{\max}\{P_{t+1,i}^{-1}\}\nonumber\\
=&\max_{i=1,\dots,n}\lambda_{\max}\bigg\{\sum_{j=1}^{n}a_{ji}(P_{t,j}^{-1}+\bm{\varphi}_{t,j}\bm{\varphi}_{t,j}^{\top})\bigg\}\nonumber\\
\leq &\max_{1\leq i\leq n}\sum_{j=1}^{n}a_{ji}\Big(\lambda_{\max}\{P_{t,j}^{-1}\}+\lambda_{\max}\{\bm{\varphi}_{t,j}\bm{\varphi}_{t,j}^{\top}\}\Big)\nonumber\\
\leq & \max_{1\leq i\leq n}\lambda_{\max}\{P_{t,i}^{-1}\}+\sum_{j=1}^{n}\|\bm{\varphi}_{t,j}\|^{2}\nonumber\\
\leq & \max_{1\leq i\leq n}\lambda_{\max}\{P_{0,i}^{-1}\}+\sum_{j=1}^{n}\sum_{k=0}^{t}\|\bm{\varphi}_{k,j}\|^{2}\nonumber\\
\leq & \lambda_{\max}\{\bm{P}_{0}^{-1}\}+\sum_{j=1}^{n}\sum_{k=0}^{t}\|\bm{\varphi}_{k,j}\|^{2}.
\end{align}
From (\ref{Pt}) and the connection between determinant and eigenvalues of a matrix, it is easy to conclude that 
$$
\log(\vert\bm{P}_{t+1}^{-1}\vert)\leq mn\log\Big(\max_{1\leq i\leq n}\lambda_{\max}\{P_{t+1,i}^{-1}\}\Big)\leq mn\log (r_{t}).
$$ 
Consequently, \emph{Theorem \ref{thm1}} follows from this and \emph{Lemma \ref{lem4}} immediately.

\subsection{Proof of \emph{Theorem \ref{thm2}}}

By the definition of $\bm{b}_{k}$ in (\ref{7}), we know that 
$$
\bm{\Phi}_{k}\bm{\Phi}_{k}^{\top}=\bm{\Phi}_{k}\bm{b}_{k}\bm{\Phi}_{k}^{\top}+\bm{\Phi}_{k}(\bm{b}_{k}\bm{\Phi}_{k}^{\top}\bm{P}_{k}\bm{\Phi}_{k})\bm{\Phi}_{k}^{\top}. 
$$
Then by noticing that $\bm{b}_{k}, \bm{\Phi}_{k}$ and $\bm{P}_{k}$ are (block) diagonal matrices, and $\bm{\Phi}_{k}^{\top}\bm{P}_{k}\bm{\Phi}_{k}=O(1), \ \hbox{a.s.}$, we know that
\begin{align}\label{key}
&\sum_{i=1}^{n}\sum_{k=0}^{t}R_{k,i}=\sum_{i=1}^{n}\sum_{k=0}^{t}(\bm{\varphi}_{k,i}^{\top}\widetilde{\bm{\theta}}_{k,i})^{2}=\sum_{k=0}^{t}\widetilde{\bm{\Theta}}_{k}^{\top}\bm{\Phi}_{k}\bm{\Phi}_{k}^{\top}\widetilde{\bm{\Theta}}_{k}\nonumber\\
=&\sum_{k=0}^{t}\widetilde{\bm{\Theta}}_{k}^{\top}\bm{\Phi}_{k}\bm{b}_{k}\bm{\Phi}_{k}^{\top}\widetilde{\bm{\Theta}}_{k}+\sum_{k=0}^{t}\widetilde{\bm{\Theta}}_{k}^{\top}\bm{\Phi}_{k}(\bm{b}_{k}\bm{\Phi}_{k}^{\top}\bm{P}_{k}\bm{\Phi}_{k})\bm{\Phi}_{k}^{\top}\widetilde{\bm{\Theta}}_{k}\nonumber\\
=&O\bigg(\sum_{k=0}^{t}\widetilde{\bm{\Theta}}_{k}^{\top}\bm{\Phi}_{k}\bm{b}_{k}\bm{\Phi}_{k}^{\top}\widetilde{\bm{\Theta}}_{k}\bigg).
\end{align}
Substituting this into \emph{Theorem \ref{thm1}} $1)$, we conclude that (\ref{Coro}) holds.

\subsection{Proof of \emph{Theorem \ref{thm3}}}

For ease of representation, let $a_{ij}^{(s)}$ be the $(i,j)$-th entry of the matrix $\mathcal{A}^{s}, s\geq 1$. Note that $a_{ij}^{(1)}=a_{ij}$. By \emph{Assumption \ref{asm3}} and \emph{Remark 3.1}, we know that $a_{ji}^{(D_{\mathcal{G}})}\geq a_{\min}>0$, where $a_{\min}=\min\limits_{i,j\in\mathcal{V}}a_{ij}^{(D_{\mathcal{G}})}>0$, and $D_{\mathcal{G}}$ is the diameter of the graph $\mathcal{G}$. Consequently, it is not difficult to see that for any $k>D_{\mathcal{G}}$,  $a_{ji}^{(k)}\geq a_{\min}$ holds.

By (\ref{7}), it is easy to see that for any $t\geq 0$, 
$$
\begin{aligned}
\text{vec}\{\bm{P}_{t+1}^{-1}\}=&\mathscr{A}\text{vec}\{\bar{\bm{P}}_{t+1}^{-1}\}=\mathscr{A}\text{vec}\{\bm{P}_{t}^{-1}\}+\mathscr{A}\text{vec}\{\bm{\Phi}_{t}\bm{\Phi}_{t}^{\top}\}\\
=&\mathscr{A}^{t+1}\text{vec}\{\bm{P}_{0}^{-1}\}+\sum_{k=0}^{t}\mathscr{A}^{t-k+1}\text{vec}\{\bm{\Phi}_{k}\bm{\Phi}_{k}^{\top}\},
\end{aligned}
$$ 
which implies that for any $t\geq D_{\mathcal{G}}$,
\begin{align}
P_{t+1,i}^{-1}= &\sum_{j=1}^{n}a_{ji}^{(t+1)}P_{0,j}^{-1}+\sum_{j=1}^{n}\sum_{k=0}^{t}a_{ji}^{(t-k+1)}\bm{\varphi}_{k,j}\bm{\varphi}_{k,j}^{\top}\nonumber\\
\geq &\sum_{j=1}^{n}a_{ji}^{(t+1)}P_{0,j}^{-1}+\sum_{j=1}^{n}\sum_{k=0}^{t-D_{\mathcal{G}}+1}a_{ji}^{(t-k+1)}\bm{\varphi}_{k,j}\bm{\varphi}_{k,j}^{\top}\nonumber\\
\geq &a_{\min}\sum_{j=1}^{n}P_{0,j}^{-1}+a_{\min}\sum_{j=1}^{n}\sum_{k=0}^{t-D_{\mathcal{G}}+1}\bm{\varphi}_{k,j}\bm{\varphi}_{k,j}^{\top}.
\end{align}
Then we have 
$$
\lambda_{\min}\{\bm{P}_{t+1}^{-1}\}\geq a_{\min}\lambda_{\min}\bigg\{\sum_{j=1}^{n}P_{0,j}^{-1}+\sum_{j=1}^{n}\sum_{k=0}^{t-D_{\mathcal{G}}+1}\bm{\varphi}_{k,j}\bm{\varphi}_{k,j}^{\top}\bigg\}. 
$$
Note also that 
$$
\|\widetilde{\bm{\Theta}}_{t+1}\|^{2}\leq\widetilde{\bm{\Theta}}_{t+1}^{\top}\bigg[\frac{\bm{P}_{t+1}^{-1}}{\lambda_{\min}\{\bm{P}_{t+1}^{-1}\}}\bigg]\widetilde{\bm{\Theta}}_{t+1}.
$$ 
Hence, by $2)$ in \emph{Theorem \ref{thm1}}, we know that \emph{Theorem \ref{thm3}} holds.

\section{Concluding remarks}

In this paper, we have established a convergence theory for a basic class of distributed LS algorithms, under quite general conditions on the measured information or data used in the estimation. The accumulated regret of adaptive predictors has been shown to have a celebrated logarithm increase without any excitation condition imposed on the system data, and the convergence rate of the distributed LS estimates has also been established under a cooperative excitation condition, which can be regarded as an extension of the weakest possible excitation condition known for the convergence of the classical LS. Neither independence and stationarity, nor Gaussian property, are required in our results. Moreover, the cooperative excitation condition introduced and used in the paper indicates that the distributed LS can fulfill the estimation task cooperatively, even if any individual sensor cannot due to lack of necessary excitation.


\begin{thebibliography}{1}

\bibitem{Sayed2013}
A. H. Sayed, S. Y. Tu, J. Chen, X. Zhao, and Z. J. Towfic, ``Diffusion strategies for adaptation and learning over networks: an examination of distributed strategies and network behavior,'' \emph{IEEE Signal Processing magazine}, vol. 30, no. 3, pp. 155--171, May 2013.

\bibitem{Sayed2014b}
A. H. Sayed, ``Adaptive networks,'' \emph{Proceedings of the IEEE}, vol. 102, no. 4, pp. 460--497, April 2014.

\bibitem{Xie2018Auto}
S. Y. Xie and L. Guo, ``A necessary and sufficient condition for stability of LMS-based consensus adaptive filters,'' \emph{Automatica}, vol. 93, pp. 12--19, July 2018.

\bibitem{Xie2018SIAM}
S. Y. Xie and L. Guo, ``Analysis of normalized least mean squares-based consensus adaptive filters under a general information condition,'' \emph{SIAM Journal on Control and Optimization}, vol. 56, no. 5, pp. 3404--3431, Sept. 2018.

\bibitem{Xie2018TAC}
S. Y. Xie and L. Guo, ``Analysis of distributed adaptive filters based on diffusion strategies over sensor networks'', \emph{IEEE Transactions on Automatic Control}, vol. 63, no. 11, pp. 3643--3658, Nov. 2018.

\bibitem{Har2019}
I. E. K. Harrane, R. Flamary and C. Richard, ``On reducing the communication cost of the diffusion LMS algorithm,'' \emph{IEEE Transactions on Signal and Information Processing over Networks}, vol. 5, no. 1, pp. 100--112, March 2019.

\bibitem{Sayed2006}
A. H. Sayed and C. G. Lopes, ``Distributed recursive least-squares strategies over adaptive networks,'' in \emph{40th Asilomar Conference on Signals, Systems and Computers}, Pacific, Grove, CA, Oct. 2006, pp. 233--237.


\bibitem{Mateos2012}
G. Mateos and G. B. Giannakis, ``Distributed recursive least-squares: stability and performance analysis,'' \emph{IEEE Transactions on Signal Processing}, vol. 60, no. 7, pp. 3740--3754, July 2012.

\bibitem{Xiao2006}
L. Xiao, S. Boyd, and S. Lall, ``A space-time diffusion scheme for peer-to-peer least-squares estimation,'' in \emph{Proceedings of 5th International Conference on Information Processing in Sensor Networks (IPSN 2006)}, Nashville, TN, April 2006, pp. 168--176.

\bibitem{Cattivelli2008}
F. S. Cattivelli, C. G. Lopes, and A. H. Sayed, ``Diffusion recursive least squares for distributed estimation over adaptive networks,'' \emph{IEEE Transactions on Signal Processing}, vol. 56, no. 5, pp. 1865--1877, May 2008.

\bibitem{Bertrand2011}
A. Bertrand, M. Moonen, and A. H. Sayed, ``Diffusion bias-compensated RLS estimation over adaptive networks,'' \emph{IEEE Transactions on Signal Processing}, vol. 59, no. 11, pp. 5212--5224, Nov. 2011.

\bibitem{Ara2014}
R. Arablouei, K. Dogancay, S. Werner, and Y-F Huang, ``Adaptive distributed estimation based on recursive least-squares and partial diffusion,'' \emph{IEEE Transactions on Signal Processing}, vol. 62, no. 14, pp. 3510--3522, July 2014.

\bibitem{Vah2017}
V. Vahidpour, A. Rastegarnia, A. Khalili and S. Sanei, ``Analysis of partial diffusion recursive least squares adaptation over noisy links,'' \emph{IET Signal Processing}, vol. 11, no. 6, pp. 749--757, August 2017.

\bibitem{Ras2019}
A. Rastegarnia, ``Reduced-communication diffusion RLS for distributed estimation over multi-agent networks,'' \emph{IEEE Transactions on Circuits and Systems II: Express Briefs}, doi: 10.1109/TCSII.2019.2899194.

\bibitem{Yu2019}
Y. Yu, H. Zhao, R. C. de Lamare, Y. Zakharov and L. Lu, ``Robust distributed diffusion recursive least squares algorithms with side information for adaptive networks," \emph{IEEE Transactions on Signal Processing}, vol. 67, no. 6, pp. 1566--1581, March 2019.


\bibitem{Carli2008}
R. Carli, A. Chiuso, L. Schenato, and S. Zampieri, ``Distributed Kalman filtering based on consensus strategies,'' \emph{IEEE Journal on Selected Areas in Communications}, vol. 26, no. 4, pp. 622--633, May 2008.

\bibitem{Bat2015}
G. Battistelli, L. Chisci, G. Mugnai, A. Farina, and A. Graziano, ``Consensus-based linear and nonlinear filtering,'' \emph{IEEE Transactions on Automatic Control}, vol. 60, no. 5, pp. 1410--1415, May 2015.

\bibitem{Liu2018}
Q. Liu , Z. Wang, X. He, and D. H. Zhou, ``On Kalman-consensus filtering with random link failures over sensor networks,'' \emph{IEEE Transactions on Automatic Control}, vol. 63, no. 8, pp. 2701--2708, Aug. 2018.

\bibitem{Das2017}
S. Das and J. Moura, ``Consensus+innovations distributed Kalman filter with optimized gains,'' \emph{IEEE Transactions on Signal Processing}, vol. 65, no. 2, pp. 467--481, Jan. 2017.

\bibitem{Astrom1973}
K. J. $\mathring{\text{A}}$str\" om and B. Wittenmark, ``On self tuning regulators,'' \emph{Automatica}, vol. 9, no. 2, pp. 185--199, March 1973.

\bibitem{Guo1995}
L.~Guo, ``Convergence and logarithm laws of self-tuning regulators,'' \emph{Automatica}, vol. 31, no. 3, pp. 435--450, March 1995.

\bibitem{Ljung1976}
L. Ljung, ``Consistency of the least-squares identification method,'' \emph{IEEE Transactions on Automatic Control}, vol. 21, no. 5, pp. 779--781, Oct. 1976.

\bibitem{Moore1978}
J. B. Moore, ``On strong consistency of least squares identification algorithm,'' \emph{Automatica}, vol. 14, no. 5, pp. 505--509, Sep. 1978.

\bibitem{Chen1982}
H. F. Chen, ``Strong consistency and convergence rate of least squares identification,'' \emph{Science in China Series A - Mathematics, Physics, Astronomy \& Technological Science}, vol. 25, no. 7, pp. 771--784, 1982.

\bibitem{Lai1982}
T. L. Lai and C. Z. Wei, ``Least squares estimates in stochastic regression models with applications to identification and control dynamic systems,'' \emph{Annals of Statistics}, vol. 10, no. 1, pp. 154--166, March 1982.

\bibitem{Lai1986}
T. L. Lai and C. Z. Wei, ``Extended least squares and their applications to adaptive control and prediction in linear systems,'' \emph{IEEE Transactions on Automatic Control}, vol. 31, no. 10, pp. 898--906, Oct. 1986.

\bibitem{Chen1986}
H. F. Chen and L. Guo, ``Convergence rate of least-squares identification and adaptive control for stochastic systems,'' \emph{International Journal of Control}, vol. 44, no. 5, pp. 1459--1476, Nov. 1986.

\bibitem{Chen1991}
H. F. Chen and L. Guo, \emph{Identification and Stochastic Adaptive Control}, Birkh\"{a}suser, Boston, 1991.

\bibitem{Guo1991}
L. Guo and H. F. Chen,  ``The $\mathring{\text{A}}$str\" om-Wittenmark self-tuning regulator revised and ELS-based adaptive tracker,'' \emph{IEEE Transactions on Automatic Control}, vol. 36, no. 7, pp. 802--812, July 1991.

\bibitem{Ljung1977}
L. Ljung, ``Analysis of recursive stochastic algorithms,'' \emph{IEEE Transactions on Automatic Control}, vol. 22, no. 4, pp. 551--575, Aug. 1977.

\bibitem{Goodwin1981}
G. C. Goodwin, P. J. Ramadge, and P. E. Caines, ``Discrete time stochastic adaptive control,'' \emph{SIAM Journal on Control and Optimization}, vol. 19, no. 6, pp. 829--853, 1981.

\bibitem{Kumar1990}
P. R. Kumar, ``Convergence of adaptive control schemes using least-squares parameter estimates,'' \emph{IEEE Transactions on Automatic Control}, vol. 35, no. 4, pp. 416--424, April 1990.

\bibitem{Julier1997}
S. Julier and J. Uhlmann, ``A non-divergent estimation algorithm in the presence of unknown correlations,'' in \emph{Proceedings of the 1997 American Control Conference}, Albuquerque, NM, USA, April 1997, pp. 2369--2373.

\bibitem{Chen2002}
L. Chen, P. Arambel, and P. Mehra, ``Estimation under unknown correlation: covariance intersection revisited,''  \emph{IEEE Transactions on Automatic Control}, vol. 47, no. 11, pp. 1879--1882, May 2002.

\bibitem{Chow1978}
Y. S.~Chow and H.~Teicher, \emph{Probability Theory}, New York: Springer, March 1978.

\bibitem{Guo}
L. Guo, \emph{Time-Varying Stochastic Systems--Stability and Adaptive Theory (Second Edition)}, Science Press, 2020.

\bibitem{Zhong2010}
M. Zhong and C. G. Cassandras, ``Asynchronous distributed optimization with event-driven communication,'' \emph{IEEE Transactions on Automatic Control}, vol. 55, no. 12, pp. 2735--2750, Dec. 2010.

\bibitem{Xie2020}
S. Y. Xie and L. Guo, ``Analysis of compressed distributed adaptive filters,'' \emph{Automatica}, vol. 112, 108707, Feb. 2020.

\bibitem{Godsil2014}
C. Godsil and G. Royle, \emph{Algebraic Graph Theory}, Springer-Verlag, 2014.

\bibitem{Hos2016}
S. Hosseini, A. Chapman, and M. Mesbahi, ``Online distributed convex optimization on dynamic networks,'' \emph{IEEE Transactions on Automatic Control}, vol. 61, no. 11, pp. 3545--3550, Nov. 2016.

\bibitem{Sha2018}
S. Shahrampour and A. Jadbabaie, ``Distributed online optimization in dynamic environments using mirror descent,'' \emph{IEEE Transactions on Automatic Control}, vol. 63, no. 3, pp. 714--725, March 2018.

\bibitem{Lai1986A}
T. L. Lai, ``Asymptotically efficient adaptive control in stochastic regression models,'' \emph{Advances in Applied Mathematics}, vol. 7, no. 1, pp. 23--45, March 1986.

\bibitem{Ky1950}
K. Fan, ``On a theorem of Weyl concerning eigenvalues of linear transformations,'' \emph{Proceedings of the National Academy of Sciences of the United States of America}, vol. 36, no. 1, pp. 31--35, Jan. 1950.

\end{thebibliography}
\end{document}